\documentclass[aps,pra,twocolumn,groupedaddress]{revtex4-1}
\usepackage{color}
\usepackage{amsmath,amsfonts,amssymb}
\usepackage{amsthm}
\usepackage{bm}
\usepackage{graphicx}
\usepackage{subfigure}
\usepackage{bbm}
\usepackage{epstopdf}
\usepackage{epsfig}
\usepackage{verbatim}
\usepackage{array}
\usepackage{ulem}
\usepackage{notes2bib}
\usepackage[T1]{fontenc}
\usepackage{tgtermes}

\usepackage{dcolumn}
\usepackage{physics}
\usepackage{braket}
\usepackage[unicode=true,bookmarks=true,bookmarksnumbered=false,bookmarksopen=false, breaklinks=true,pdfborder={0 0 1},backref=false,colorlinks=true]{hyperref}
\hypersetup{linkcolor=magenta, urlcolor=blue, citecolor=blue, pdfstartview={FitH}, unicode=true}

\begin{document}

\title{Accelerated quantum Monte Carlo simulations of the attractive Hubbard model on the kagome lattice}

\author{Jie Zhang}
\affiliation{School of Physics and Technology, Wuhan University, Wuhan 430072, China}
\author{Xiang Li}
\affiliation{School of Physics and Technology, Wuhan University, Wuhan 430072, China}
\author{Yu Wang}
\email{yu.wang@whu.edu.cn}
\affiliation{School of Physics and Technology, Wuhan University, Wuhan 430072, China}

\date{\today}

\begin{abstract}
The recent discovery of several families of kagome materials and experimental realization of optical kagome lattices have stimulated growing numerical studies of interaction-driven correlated states on the kagome lattice. Among the available numerical approaches, determinant quantum Monte Carlo (DQMC) is a powerful method for investigating such strongly correlated states. However, the accessible system sizes of existing DQMC simulations remain limited, preventing reliable finite-size scaling analyses. Here we develop a general acceleration scheme based on fast Fourier transform (FFT) for propagator multiplications on composite lattices and combine it with the delay-update algorithm, enabling simulations on system sizes twice as large as those of previous DQMC studies, allowing reliable finite-size scaling analyses of the attractive kagome-lattice Hubbard model. Our large-scale simulations reveal the interaction-driven zero-temperature superfluid quantum criticality at the Dirac filling and provide reliable estimates of the associated critical exponents. Besides, we find no evidence that the previously proposed triangle-rule charge-density-wave order survives in the thermodynamic limit, suggesting that it is likely a finite-size effect. Moreover, for system sizes accessible in current two-dimensional optical lattice experiments, the combined FFT and delay-update scheme exhibits an effective computational cost scaling as $N^{2.49}$, substantially below the $\mathcal{O}(N^3)$ computational cost of conventional DQMC simulations.  
\end{abstract}

\maketitle

\section{Introduction}
In recent years, the discovery of kagome materials, most notably the $\mathrm{AV}_3\mathrm{Sb}_5$ (A= K, Rb, Cs) family~\cite{NatureYe2018,NaturePhysLiu2018,Ortiz2019}, together with the realization of optical kagome lattices~\cite{Jooptical2012,Schaferoptical2020}, has renewed broad interests in correlated physics on the kagome lattice. Its unique lattice geometry and band structure, featuring geometrical frustration, flat band, Dirac fermions, and van Hove singularities, give rise to a rich variety of interaction-driven quantum phases~\cite{Mekata2003}. Among these phases, superfluid (or superconducting) and charge-density-wave (CDW) orders have attracted particular attention because the interplay between geometric frustration, filling-dependent band structure, and interactions gives rise to a delicate competition between the two orders~\cite{CDWSFKang2022,CDWSFJiang2021,CDWSFFeng2021,CDWSFMielke2022,CDWSFNie2022,CDWSFChen2021,CDWSFYu2021,CDWSFWu2021,CDWSFTan2021,CDWSFZhao2021,CDWSFScammell2023}. At present, the nature of this competition remains under active debate. A natural starting point for understanding this competition is to investigate how simple interacting models describe these competing orders. Among such simple models, the attractive Hubbard model provides a prototypical platform. Previous determinant quantum Monte Carlo (DQMC) studies of this model have revealed superfluid critical behavior at selected fillings and suggested a possible triangle-rule CDW tendency at the Dirac filling~\cite{QMCGuo2023}. However, owing to the high computational cost of DQMC simulations on the kagome-lattice Hubbard model, the accessible system sizes remain limited, making reliable finite-size scalings challenging. This limitation motivates the development of efficient techniques to accelerate DQMC simulations, thereby enabling simulations on substantially larger kagome lattices.

In conventional implementations of DQMC simulations for Hubbard-type models, the computational cost is dominated by two operations: Monte Carlo updates of the auxiliary field and repeated multiplications of imaginary-time propagators~\cite{Blankenbecler1981,Hirsch1983}. The former consists of a large number of low-arithmetic-intensity rank-one Green's-function updates, whereas the latter requires repeated multiplications of dense propagator matrices. Delayed-update algorithms have substantially accelerated auxiliary-field updates by accumulating multiple rank-one Green's-function updates into a single BLAS-3 matrix update, thereby achieving significant speedups through increased arithmetic intensity~\cite{delayAlvarez2008,delayNukala2009,delaySun2024,delaySun2025,delayDu2025}. For propagator multiplications, several distinct acceleration strategies have been developed. Checkerboard decomposition factorizes the dense kinetic propagator into a sequence of sparse bond propagators at the expense of additional Trotter errors and lattice-dependent bond decompositions~\cite{checkerChuang2012,checkerLee2013,checkerCohenStead2024}. Low-rank approaches evaluate the matrix exponential through Krylov-subspace projections, but are generally effective only in the low-density limit~\cite{lowrankHe2019,lowrankCelledoni2000}. Fast Fourier transform (FFT) method exploits the diagonal representation of the kinetic propagator in momentum space, reducing propagator multiplications to Fourier transforms and diagonal matrix operations, and this method has proven highly effective for square and cubic lattices~\cite{FFTBai2009,FFTSong2025}, while its extension to composite lattices requires a more general formalism.       

In this work, we generalize the FFT method to composite lattices, where propagator multiplications are reduced to Fourier transforms and block-diagonal matrix operations, thereby reducing the computational complexity of propagator multiplications from $\mathcal{O}(N^3)$ to $\mathcal{O}(N^2 \log(N))$. Combined with the delay-update scheme~\cite{delayAlvarez2008,delayNukala2009,delaySun2024,delaySun2025,delayDu2025}, our approach enables simulations on system sizes twice as large as those accessible in previous studies~\cite{QMCGuo2023}. This allows reliable finite-size scaling analyses of both the superfluid (SF) and the CDW orders. We focus on the ground-state properties of the two orders, and employ the projector determinant quantum Monte Carlo simulations, the zero-temperature variant of DQMC simulations throughout this work.

The rest of this paper is organized as follows. In Sec.~\ref{sec:model}, the model Hamiltonian and the basic formalism of FFT-acclerated propagator multiplications are introduced. In Sec.~\ref{sec:ComputationalPerformance}, we compare the performance of FFT-acclerated propagator multiplications and delay updates in different system sizes, and analyse how the computational cost scales with system sizes. In Sec.~\ref{sec:SF}, we present large-scale simulations of the SF critical behavior. In Sec.~\ref{sec:CDW}, we investigate the stability of two types of CDW orders by performing finite-size scaling analyses on substantially larger system sizes. The conclusions and discussions are presented in Sec.~\ref{sec:conclusion}.   

\section{Model and Method}
\label{sec:model}
We consider the Hubbard model on composite lattices and use the attractive Hubbard model on the kagome lattice as a representative example to demonstrate the FFT-accelerated scheme. The Hamiltonian is given by 
\begin{equation}
\label{Hamiltonian}
H=\sum_{ ij,\sigma}(c^\dagger_{i\sigma}K_{ij} c_{j\sigma})+ U\sum_{i}n_{i\uparrow}n_{i\downarrow},
\end{equation}
in which $c^\dagger_{i\sigma}$ ($c_{i\sigma}$) is the creation (annihilation) operator on site $i$ with spin $\sigma$; $K$ is the spinless kinetic matrix, with $K_{ij}=-t$ for nearest-neighbor sites and $0$ otherwise; $n_{i\sigma}$ is the partical number operator on site $i$ with spin $\sigma$; $U<0$ is the Hubbard attraction; the hopping amplitude $t$ is set as the unit of energy for the system. In our simulations, the filling number is controlled by the trial wave function, which is constructed by occupying the lowest $N_\mathrm{occ}$ eigenstates of $K$. The corresponding total particle filling is $\rho=\frac{2N_\mathrm{occ}}{N}$, where $N$ is the total number of lattice sites. 

To enable the FFT-accelerated matrix multiplication, we Fourier transform $K$ into its momentum-space representation $\Lambda$. For a d-dimensional composite lattice with $L^d$ unit cells and $n$ sublattices per unit cell, $K$ has dimension $N=nL^d$, and is Fourier transformed as:
\begin{equation}\label{eq:condition}
  \left[\qty(F^{\dagger})^{\otimes d}\otimes I_n\right] K
  \left[\vphantom{{(F^{\dagger})}^{\otimes d}} F^{\otimes d}\otimes I_n\right]
  = \Lambda,
\end{equation}
in which $F$ is the one-dimensional discrete Fourier transform matrix with dimension $L$, $\otimes$ denotes the Kronecker product, and $I_n$ is the $n-$dimensional identity matrix. The resulting matrix $\Lambda$ is block-diagonal, consisting of $L^d$ independent $n-$dimensional blocks.

To illustrate this general framework, we now consider the spinless kinetic matrix $K$ on the kagome lattice, which can be expressed as
\begin{align}\label{eq:kag_kron_plus}
  K=&I_L\otimes\qty(I_L+K_s)\otimes \delta_a + \qty(I_L+K_s)\otimes I_L\otimes \delta_b\notag\\
  &+\qty(I_L\otimes I_L+K_s\otimes K_s^{T})\otimes \delta_c + \text{h.c.}
\end{align}
In Eq.~(\ref{eq:kag_kron_plus}), $I_L$ is the $L$-dimensional identity matrix. The $L$-dimensional circulant matrix $K_s$ represents one-step translation along a lattice direction, with nonzero entries only on the first subdiagonal and the top-right corner. The 3-dimensional matrices $\delta_{a,b,c}$ encode the intra-cell hopping, each containing a single nonzero entry at $(\delta_a)_{2,1}$, $(\delta_b)_{3,1}$ and $(\delta_c)_{3,2}$, respectively.
Applying the Fourier transform to Eq.~(\ref{eq:kag_kron_plus}), only $K_s$ is diagonalized as $F K_s F^\dagger = \Lambda_s$, while other matrices remain unchanged. The spinless kinetic matrix $K$ therefore becomes
\begin{align}\label{eq:kag_block}
  K=&\qty(F\otimes F\otimes I_3)[I_L \otimes\qty(I_L+\Lambda_s)\otimes \delta_a+\qty(I_L+\Lambda_s)\otimes I_L\otimes \notag\\
  &\delta_b+\qty(I_L\otimes I_L+\Lambda_s\otimes \Lambda_s^*)\otimes \delta_c+ \text{h.c.}]\qty(F^\dagger\otimes F^\dagger\otimes I_3)\notag\\
  =&\qty(F\otimes F\otimes I_3)\Lambda_{\text{kag}}^{\qty{3\times 3}}\qty(F^\dagger\otimes F^\dagger\otimes I_3),
\end{align}
where $\Lambda_{\text{kag}}^{\qty{3\times 3}}$ is the 3$L^2$-dimensional matrix composed of $L^2$ independent $3 \times 3$ diagonal blocks. The same procedure applies directly to other lattices, whose block-diagonal forms are summarized in Appendix~\ref{append:defofmag}. In general, the spinless kinetic matrix $K$ on a general composite lattice can be block-diagonalized into $\Lambda^{\qty{n\times n}}$, which is an $nL^d$-dimensional matrix composed of $L^d$ independent $n\times n$ diagonal blocks. 

In DQMC simulations, the kinetic propagator is repeatedly multiplied by dense matrices arising from the imaginary-time evolution. Therefore, one of the computational bottlenecks is the matrix-matrix multiplication
\begin{equation}
\label{densemul}
e^{-\Delta_{\tau} K}M,
\end{equation}
in which $\Delta_{\tau}$ is the imaginary time step and $M$ is the $N$-dimensional dense matrix. Using the block-diagonal form of $K$, this multiplication reduces to independent operations in each block, thereby significantly accelerating the matrix-matrix multiplication. For clarity, the matrix-matrix multiplication is described by a sequence of independent matrix-vector multiplications.
Since the multiplication acts independently on each column of $M$, it is sufficient to consider the multiplication between $e^{-\Delta_{\tau} K}$ and an arbitrary column vector $\vb{v}_j$ of $M$, which can be expressed as
\begin{equation}
\label{multiplycore}
  e^{-\Delta\tau K} \vb{v}_j
  =\left[\vphantom{{(F^{\dagger})}^{\otimes d}} F^{\otimes d}\otimes I_n\right]e^{-\Delta\tau \Lambda^{\qty{n\times n}}}\left[\qty(F^{\dagger})^{\otimes d}\otimes I_n\right] \vb{v}_j.
\end{equation}
Eq.~(\ref{multiplycore}) can be interpreted as three consecutive steps. First, $\left[\qty(F^{\dagger})^{\otimes d}\otimes I_n\right] \vb{v}_j$ Fourier transforms the input vector $\vb{v}_j$ into a new vector $\vb{v}_j^{\prime}$ through FFT. Since the FFT is independently applied to the $n$ sublattice components, each containing $L^d$ entries, the computational complexity of this step is $\mathcal{O}(nL^d \log (L^d))$. Second, $\vb{v}_j^{\prime}$ is multiplied by the block-diagonal matrix $e^{-\Delta\tau \Lambda^{\qty{n\times n}}}$, which consists of $L^d$ blocks of size $n\times n$, yielding a new vector $v_j^{\prime\prime}$. The computational complexity of this step is $\mathcal{O}(nN)$. Finally, $v_j^{\prime\prime}$ is Fourier transformed back to real space through inverse FFT, with computational complexity $\mathcal{O}(nL^d \log (L^d))$. Combining the three steps above, the computational complexity of the matrix-vector multiplication is $\mathcal{O}(2nL^d \log (L^d)+nN)\approx \mathcal{O}(N \log (N))$, where the approximation holds for $N\gg n$. Consequently, the computational complexity of the matrix-matrix multiplication in Eq.~(\ref{densemul}), consisting of $N$ independent matrix-vector multiplications, is reduced from $\mathcal{O}(N^3)$ to $\mathcal{O}(N^2 \log (N))$.

The FFT-accelerated matrix multiplication described above is combined with the delay-update algorithm~\cite{delayAlvarez2008,delayNukala2009,delaySun2024,delaySun2025,delayDu2025} in our simulations. The resulting implementation is benchmarked in the following section, followed by demonstrations of its performance in large-scale simulations of SF and CDW ordering. Unless speciﬁcally stated, all simulations are performed on an Intel Xeon E5-2678 v3 platform using 20-30 CPU cores.

\section{Computational Performance of the Accelerated DQMC}
\label{sec:ComputationalPerformance}
\begin{figure}
  \includegraphics[width=0.8\linewidth]{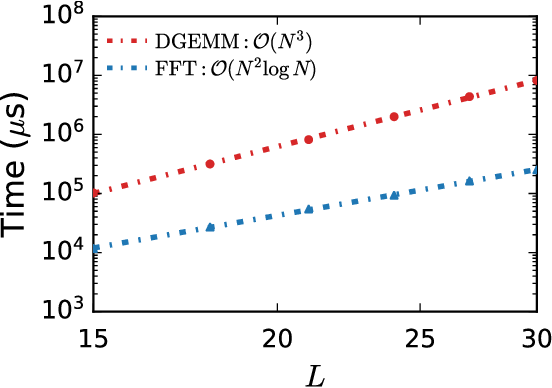}%
  \caption{\label{fig:speed_all}
  Average computational time per matrix-matrix multiplication as a function of the system size $L$ on the kagome lattice. The computational time is shown on a log-log scale at different system sizes $L$ for the standard dense matrix multiplication (DGEMM) and the FFT-accelerated matrix multiplication. Dashed lines represent linear fits.
  }
\end{figure}

To benchmark the FFT-accelerated matrix multiplication, we compare its average computational time per matrix-matrix multiplication with that of the standard dense matrix multiplication (DGEMM) on the kagome lattice, as shown in Fig.~\ref{fig:speed_all}. The FFT-accelerated matrix multiplication substantially outperforms DGEMM, and the advantage becomes increasingly pronounced as the system size $L$ increases, demonstrating the effectiveness of the proposed FFT acceleration.

\begin{figure}
  \includegraphics[width=1.0\linewidth]{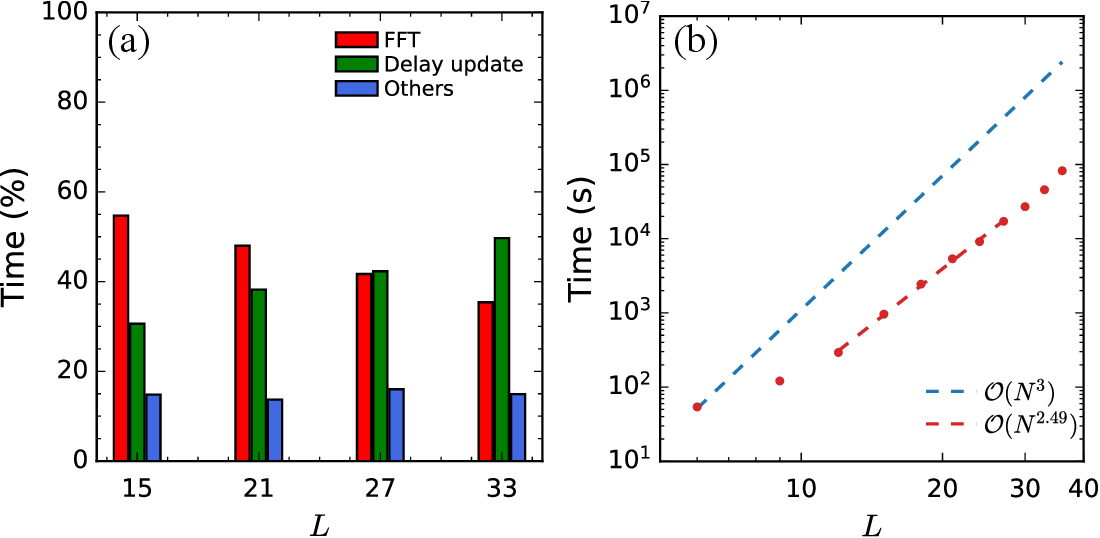}%
  \caption{\label{fig:percent}(a) Percentages of the computational time occupied by each process in our simulations at different system sizes $L$. (b) The total computational time is shown on a log-log scale for different system sizes $L$. The blue dashed line indicates the $\mathcal{O}(N^3)$ reference scaling ($N=3L^2$), while the red dashed line shows a linear fit within the range $12\leqslant L\leqslant 27$. The simulations are performed for a representative parameter set with $U=-5.0$ and $\rho=\frac{2}{3}$, using 50 warmup steps, 50 measure steps and a projection length of $\beta=10$.}
\end{figure}

Although the FFT multiplication substantially accelerates the matrix-matrix multiplication, the overall computational cost of our simulations is determined by the combined cost of all processes. Hence, it is necessary to analyse the percentage of runtime spent in each process in our simulations at different system sizes $L$. Since the major computational processes in our PQMC simulations, including FFT multiplication, delay update, and measurement all scale linearly with the number of Monte Carlo steps and projection length, their relative runtime percentages are insensitive to the total simulation length. Therefore, to access larger system sizes efficiently, the runtime analysis is performed using $50$ warmup steps, $50$ measurement steps, and a projection length of $\beta=10$, which are sufficient to provide a representative runtime percentages of different computational processes. We calculate the percentage of runtime spent in each process in our simulations for different system sizes $L$, as summarized in Fig.~\ref{fig:percent} (a). As the system size $L$ increases, the computational bottleneck gradually shifts from the FFT-accelerated matrix multiplication to the delay-update process, with the crossover occurring around $L=27$, corresponding to $N=2187$ lattice sites. This behavior originates from the competition between the $\mathcal{O}(N^2 \log(N))$ computational complexity of FFT multiplication and the small-prefactor $\mathcal{O}(N^3)$ computational complexity of delay update~\cite{delayAlvarez2008,delayNukala2009,delaySun2024,delaySun2025,delayDu2025}. 
Therefore, for large system sizes (over $L=27$ in our case), the computational complexity of our simulations is ultimately governed by the delay-update process and scales as $\mathcal{O}(N^3)$.

However, approaching extremely large system sizes is not always required for simulations targeting two-dimensional optical-lattice experiments, where reliable measurements are typically performed on systems containing hundreds to a few thousand lattice sites~\cite{2DMeng2023,2DYang2021,2DHartke2023}. For the kagome lattice considered here, this size corresponds approximately to $12\leqslant L\leqslant 27$. Therefore, it is important to characterize the computational performance in the experimentally relevant size regime. We calculate the total computational time at different system sizes $L$, as shown in Fig.~\ref{fig:percent} (b). The measured total computational time remains consistently below the $\mathcal{O}(N^3)$ reference scaling line (the blue dashed line) over the entire accessible system sizes. This hebavior is attributed to the FFT-accelerated propagator multiplication, which lowers the computational complexity of the propagator multiplication from $\mathcal{O}(N^3)$ to $\mathcal{O}(N^2 \log(N))$. More importantly, for the experimentally relevant sizes, the linear fit (the red dashed line) of $\log(\mathrm{Time})$ as a function of $\log(L)$ yields $\mathrm{Time}\sim L^{4.98}$, i.e. $\mathrm{Time}\sim N^{2.49}$.
This effective scaling behavior arises from the comparable contributions of the FFT-accelerated propagator multiplication and the delay-update process in this size regime,  and thus their competition leads to an effective scaling behavior between $\mathcal{O}(N^2 \log(N))$ and $\mathcal{O}(N^3)$.

\section{large-scale simulations of kagome superfluid} 
\label{sec:SF}
On the kagome lattice, the Dirac filling ($\rho=\frac{2}{3}$) was reported to suffer from significant finite-size effects when determining the SF critical behavior, preventing previous DQMC simulations from performing a reliable finite-size scaling analysis because of the limited accessible lattice sizes (up to $L=12$)~\cite{QMCGuo2023}. Employing our accelerated DQMC scheme, we investigate the SF critical behavior at the challenging Dirac filling on system sizes from $L=12$ to $L=24$. These substantially large system sizes enable reliable finite-size scalings required for an accurate determination of the critical behavior~\cite{Parisen2015Fermionic}.
\begin{figure}
  \includegraphics[width=1.0\linewidth]{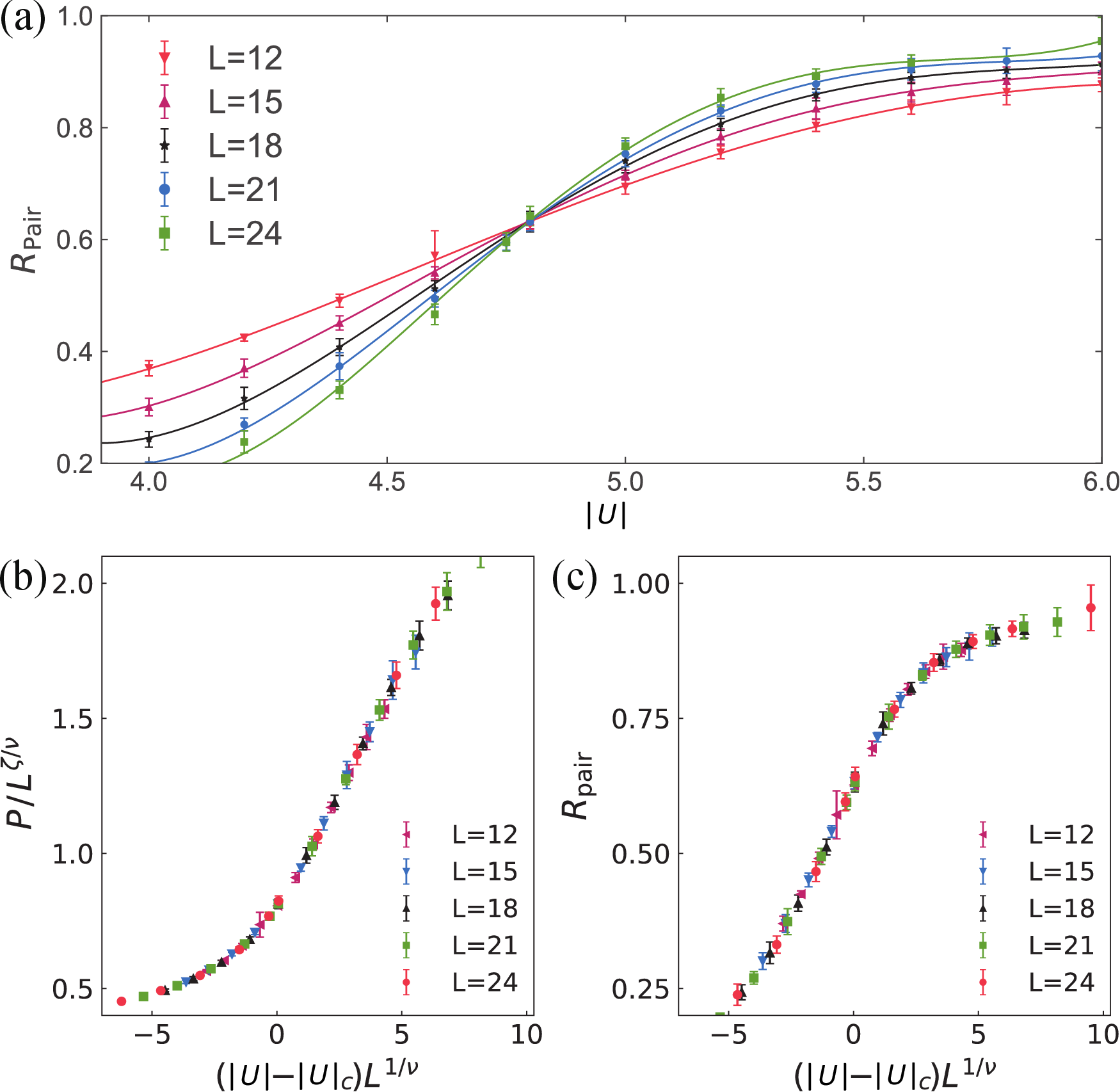}%
  \caption{\label{fig:ratio}(a) The correlation ratio $R_{\mathrm{Pair}}$ as a function of $|U|$ for different system sizes $L$ at the Dirac filling $\rho=\frac{2}{3}$. (b)-(c) Data collapses for the SF order parameter $P$ and the correlation ratio $R_{\mathrm{Pair}}$, respectively. All data points are obtained with 500 warm steps, 500 measure steps, and projection length $\beta=30$.}
\end{figure}

To precisely locate the critical interaction $|U|_{c}$ at which the s-wave pairing (i.e., SF) order appears, we first compute the dimensionless correlation ratio \cite{binder1981finite, Pujari2016}:
\begin{equation}
R_{\mathrm{Pair}} = 1 - \frac{S_{\mathrm{Pair}}(L,\Gamma + \delta k)}{S_{\mathrm{Pair}}(L,\Gamma)},
\end{equation}
in which $S_{\mathrm{Pair}}(L,\vec{Q})=\frac{1}{N}\sum_{ij}e^{i\vec{Q}\cdot(\vec{r}_i-\vec{r}_j)}\langle c_{i\uparrow}^\dagger c_{i\downarrow}^\dagger c_{j\downarrow}c_{j\uparrow}+c_{j\downarrow}c_{j\uparrow}c_{i\uparrow}^\dagger c_{i\downarrow}^\dagger\rangle$ is the pairing structure factor. $\Gamma=0$ is the zone center, and $\Gamma+$ $\delta\mathbf{k}$ represents the smallest available neighboring wave vector in the reciprocal space. $R_{\mathrm{Pair}}\rightarrow 1$ when the SF order is well developed, while $R_{\mathrm{Pair}}\rightarrow 0$ when the SF order disappears. Here, we employ the symmetric defination of the pairing structure factor $S_{\mathrm{Pair}}(L,\vec{Q})$, instead of the conventional form $S_{\mathrm{Pair}}(L,\vec{Q})=\frac{1}{N}\sum_{ij}e^{i\vec{Q}\cdot(\vec{r}_i-\vec{r}_j)}\langle c_{i\uparrow}^\dagger c_{i\downarrow}^\dagger c_{j\downarrow}c_{j\uparrow}+\mathrm{h.c.}\rangle$, following Refs.~\cite{Assaad2008QMC,QMCGuo2023,Xu2023Trion}. 
We calculate $R_{\mathrm{Pair}}$ as a function of $|U|$ for different system sizes $L$ at the Dirac filling $\rho=\frac{2}{3}$, as shown in Fig.~\ref{fig:ratio} (a). The curves for different $L$ exhibit a well-defined crossing with negligible finite-size drift, indicating the accessible lattice sizes in our simulations are sufficient for a reliable finite-size scaling analysis. The crossing provides an estimate of the critical interaction, $|U|_c\approx 4.8$.

We then determine the critical exponents of the SF phase transition. In the vicinity of the critical point, the SF order parameter 
\begin{equation}
P = \sqrt{\frac{1}{N}S_{\mathrm{Pair}}(L,\Gamma)}
\end{equation}
and the correlation ratio $R_{\mathrm{Pair}}$ obey the following scaling equations~\cite{Janke2008Monte,Parisen2015Fermionic,shao2016quantum,Xu2023Trion}:
\begin{eqnarray}
  &P(\delta |U|,L)=L^{\zeta/\nu} \tilde{P}\left(\delta |U| L^{\frac{1}{\nu}}\right)\label{eq:collapse_order}
  \\
  &R_{\mathrm{Pair}}(\delta |U|,L)=\tilde{R}_{\mathrm{Pair}} \left(\delta |U| L^{\frac{1}{\nu}}\right),\label{eq:collapse_ratio}
\end{eqnarray}
in which $\delta |U| = |U| - |U|_c$ is the distance from the critical point, while $\tilde{P}$ and $\tilde{R}_{\mathrm{Pair}}$ are universal scaling functions. The exponent $\zeta = -\nu(d + z - 2 + \eta)/2$, with the spatial dimension $d$ being 2, and the dynamical critical exponent $z$ being 1~\cite{senthil2004deconfined,senthil2004quantum,Sandvik2007,wang2019slater}. 
\begin{figure}
  \includegraphics[width=1.0\linewidth]{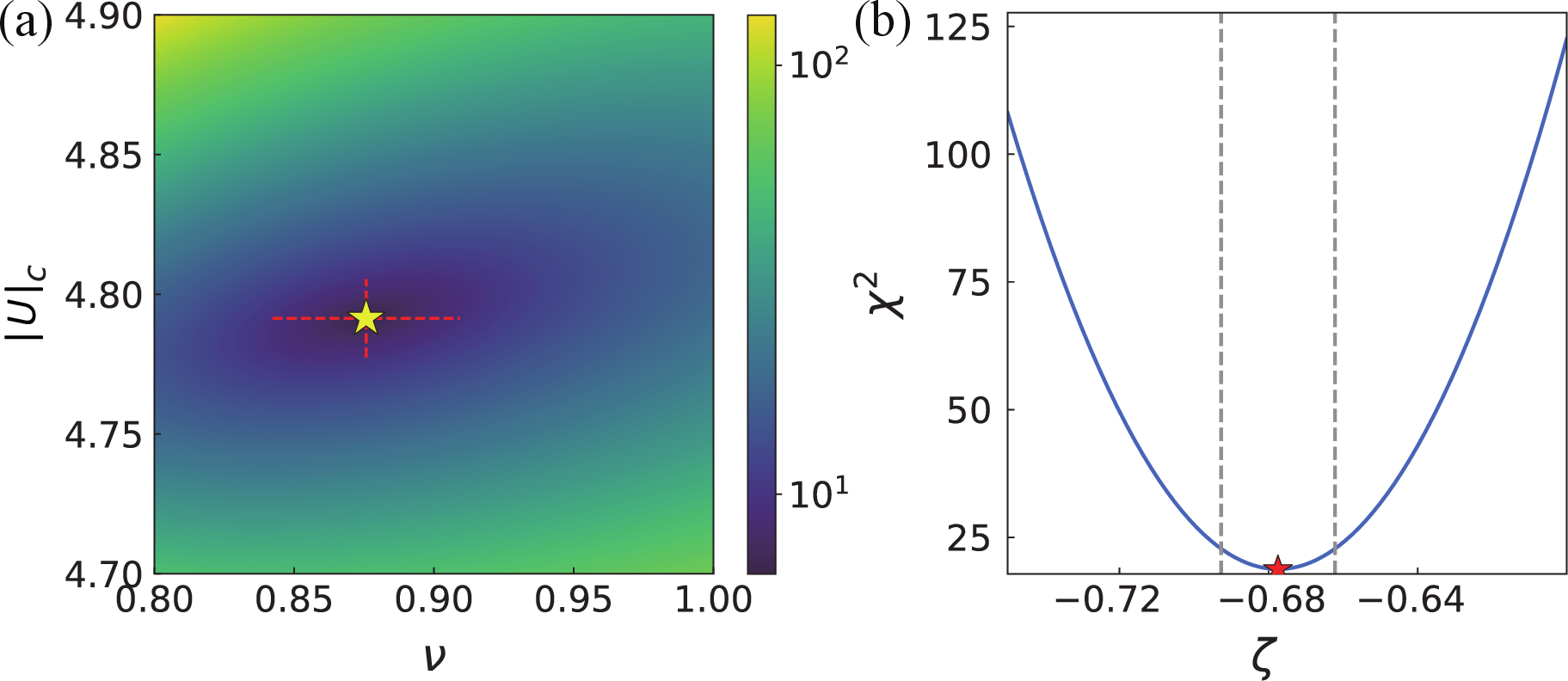}%
  \caption{\label{fig:autofit}$\chi^2$ minimization for determining the critical parameters. (a) The color map of $\chi^2$ in the ($\nu,|U|_c$) parameter plane. The yellow star marks the minimum of $\chi^2$ at $|U|_c=4.79(1)$ and $\nu=0.88(3)$. (b) Corresponding fit for the critical exponent $\zeta$, using the optimal $|U|_c$ and $\nu$ from (a). The minimum (red star) marks $\zeta=-0.68(2)$. Dashed lines in both panels represent the standard errors for the corresponding parameters.}
\end{figure}

To extract the critical exponents $\nu$ and $\eta$, we employ a systematic Finite-Size Scaling (FSS) analysis based on a $\chi^2$ minimization method \cite{melchert2009auto,Houdayer2004Low}.  First, we simultaneously determine $|U|_c$ and $\nu$ through the scaling ansatz Eq.~(\ref{eq:collapse_ratio}). We define the rescaled variable $x=\delta|U|L^{1/\nu}$, and scan a grid of trial parameter pairs $(\nu, |U|_c)$. For each trial parameter pair, the corresponding rescaled data (x, $R_{\mathrm{Pair}}$) are fitted by a fourth-order polynomial function $f(x)$. 
$\chi^2$ is then estimated as $\chi^2=\frac{1}{N_{\mathrm{DOF}}}\sum_{i}\frac{[R_{\mathrm{Pair},i}-f(x_i)]^2}{\sigma^2(R_{\mathrm{Pair},i})}$, in which $i$ runs over all rescaled data points; $\sigma(R_{\mathrm{Pair},i})$ is the statistical uncertainty of $R_{\mathrm{Pair}}$; $N_{\mathrm{DOF}}=N_{\mathrm{data}}-5$ denotes the number of degrees of freedom, with $N_{\mathrm{data}}$ being the number of data points included in the fit and 5 being the number of fitting parameters in the fourth-order polynomial.
The color map of $\chi^2$ in the ($\nu$,$|U|_c$) parameter plane is shown in Fig.~\ref{fig:autofit} (a). The optimal trial parameter pair is $|U|_c=4.79(1)$ and $\nu=0.88(3)$.
With the optimal values of $|U|_c$ and $\nu$ fixed, we then determine the critical exponent $\zeta$ through the scaling ansatz Eq.~(\ref{eq:collapse_order}).  
We scan a series of trial values of $\zeta$, and for each trial value, the corresponding rescaled data ($x$, $P/L^{\zeta/\nu}$) are fitted by a fourth-order polynomial function $f(x)$.
$\chi^2$ is then estimated as $\chi^2=\sum_{i}\frac{[P_i-L^{\zeta/\nu}f(x_i)]^2}{\sigma^2(P_i)}$, in which $\sigma(P_i)$ is the statistical uncertainty of $P$.
Fig~\ref{fig:autofit} (b) shows the calculated $\chi^2$ as a function of $\zeta$. The optimal trial value is $\zeta=-0.68(2)$. Finally, the critical exponent $\eta$ is determined by the relation $\zeta = -\nu(d + z - 2 + \eta)/2$, yielding $\eta=0.55(7)$.
Using the average values of extracted critical parameters ($|U|_c=4.79$, $\nu=0.88$, and $\zeta=-0.68$), we perform the data collapses for the SF order parameter $P$ and the correlation ratio $R_{\mathrm{Pair}}$, as shown in Fig.~\ref{fig:ratio} (b) and Fig.~\ref{fig:ratio} (c) respectively. The quality of data collapses is very high.
\begin{figure}[t]
  \includegraphics[width=0.8\linewidth]{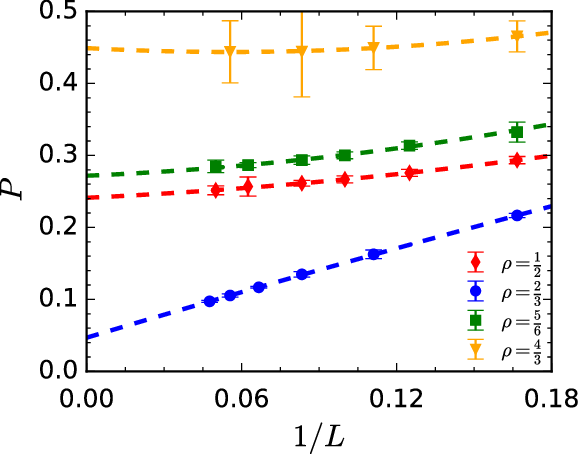}%
  \caption{\label{fig:pairingorders}Finite-size extrapolations of the pairing order parameter $P$ at $|U|=5.0$ and different fillings $\rho$. All data points are obtained with 1000 warm steps, 1000 measure steps, and projection length $\beta=30$. Quadratic fitting is applied.}
\end{figure}

Finally, we perform finite-size extrapolations of the superfluid order parameter $P$ at $|U|=5.0$ for several fillings, as shown in Fig.~\ref{fig:pairingorders}. The extrapolated values of $P$ exhibit a significant filling dependence, with the Dirac filling $\rho=\frac{2}{3}$ displaying the weakest SF order among all fillings considered.
\section{large-scale simulations of kagome CDW}
\label{sec:CDW}
Recently, two distinct CDW patterns have been proposed on the kagome lattice. 
DQMC simulations of the attractive Hubbard model on lattices up to $L=12$ at the strong coupling $|U|=8.0$ suggested that the homogeneous charge correlations may arise from configurational averaging over three triangle-rule CDW orientations~\cite{QMCGuo2023}, while the hybrid Monte Carlo study of the Holstein model up to $L=15$ identified a $\sqrt{3}\times\sqrt{3}$ CDW characterized by an ordering wave vector at the Dirac point $K$~\cite{CDWBradley2023}.
The distinct CDW patterns proposed in these studies raise the question of which ordering, if any, remains stable in the attractive Hubbard model in the thermodynamic limit. 
Addressing this question requires simulations on sufficiently large lattices.
Motivated by this, we employ our accelerated DQMC scheme to calculate the momentum-dependent CDW order
\begin{equation}
D(\vec{Q})= \sqrt{\langle \phi^{\dagger}(\vec{Q})\phi(\vec{Q})\rangle},
\end{equation}
in which $\phi(\vec{Q})=\frac{1}{N}\sum_i e^{i\vec{Q}\cdot \vec{r}_i}(n_{iA}+e^{i\frac{2\pi}{3}}n_{iB}+e^{i\frac{4\pi}{3}}n_{iC})$ is the complex CDW field. Calculating $D(\vec{Q})$ at different ordering wave vectors serves different purposes. $D(\Gamma)$ is used to test whether three triangle-rule CDW orientations exist, whereas $D(K)$ probes the possible emergence of the $\sqrt{3}\times\sqrt{3}$ CDW state. Since $D(\Gamma)$ probes the CDW intensity,  the three triangle-rule CDW orientations do not cancel under Monte Carlo averaging. Therefore, finite $D(\Gamma)$ provides direct evidence whether individual Monte Carlo configurations possess CDW patterns.  

\begin{figure}[t]
  \includegraphics[width=1.0\linewidth]{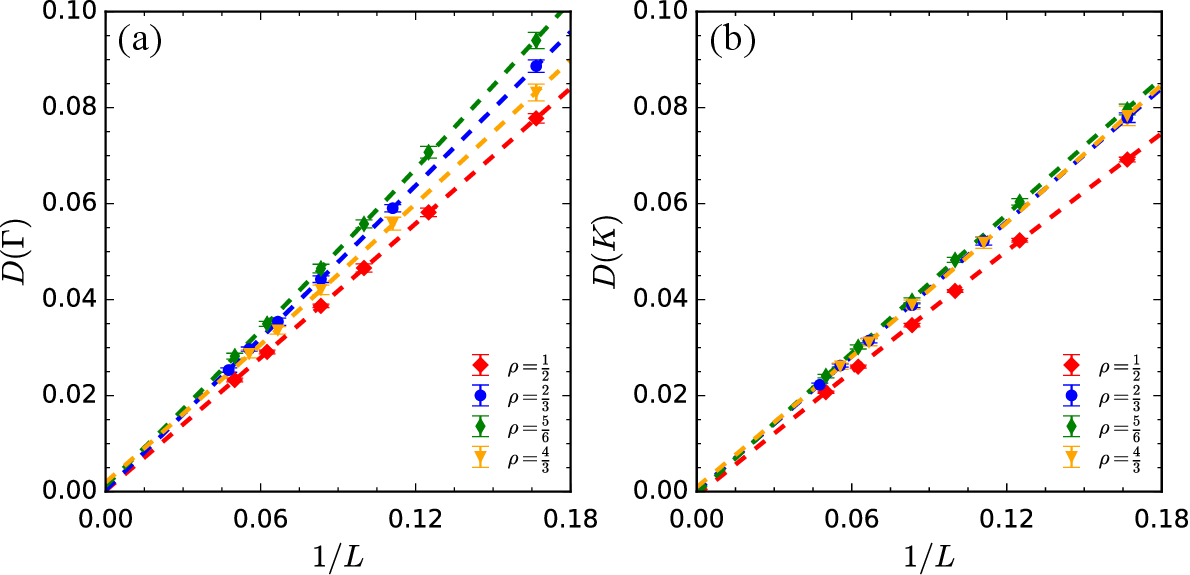}%
  \caption{\label{fig:CDWorders}Finite-size extrapolations of the CDW order parameters at the wave vector of (a) $\Gamma$ and (b) $K$ at $|U|=8.0$ for different fillings $\rho$. All data points are obtained with 1000 warm steps, 1000 measure steps, and projection length $\beta=30$. Quadratic fitting is applied.}
\end{figure}
We present the finite-size extrapolations of $D(\Gamma)$ and $D(K)$ at $|U|=8.0$ for different fillings $\rho$ in Fig.~\ref{fig:CDWorders}. For all fillings considered, $D(\Gamma)$ extrapolates to zero in the thermodynamic limit, indicating the absence of a stable triangle-rule CDW orientation in individual Monte Carlo configurations. These results suggest that the proposed triangle-rule CDW orientation in individual Monte Carlo configurations is likely a finite-size effect. Likewise, $D(K)$ also extrapolates to zero across all fillings, providing no evidence for a stable $\sqrt{3}\times \sqrt{3}$ CDW in our attractive Hubbard model.

\section{conclusion and discussion}
\label{sec:conclusion}
We develop an FFT-accelerated matrix multiplication scheme for DQMC simulations of the attractive Hubbard model on composite lattices and combine it with the delay-update algorithm~\cite{delayAlvarez2008,delayNukala2009,delaySun2024,delaySun2025,delayDu2025}, using the kagome lattice as a representative example. We demonstrate that, for system sizes accessible in two-dimensional optical-lattice experiments (hundreds to a few thousand lattice sites), the combined FFT and delay-update scheme exhibits an effective computational cost scaling as $N^{2.49}$, substantially below the $\mathcal{O}(N^3)$ computational cost of conventional DQMC simulations. The resulting improvement enables reliable finite-size scaling analyses of the superfluid transition at the Dirac filling $\rho=\frac{2}{3}$, yielding high-quality estimates for critical interaction strength of $|U|_c=4.79(1)$ and critical exponents of $\nu=0.88(3)$ and $\zeta=-0.68(2)$, overcoming the severe finite-size effects encountered in previous studies~\cite{QMCGuo2023}. Our results also suggest that the previously proposed triangle-rule CDW on the kagome lattice~\cite{QMCGuo2023} is likely a finite-size effect.

The present framework is applicable to a broad class of DQMC simulations beyond the attractive Hubbard model on the kagome lattice. Promising applications include the SU($2N$) Hubbard model~\cite{zhou2014su2n,Zhou2016SU2N,Zhou2017Thermal,Zhou2018PiFlux}, the three-component Hubbard model~\cite{Xu2023Trion,Li2023ThermalSU3,Li2024NeelCDW,Li2025TrionPiFlux}, and finite-range interacting models~\cite{delaySun2024} on composite lattices.
\acknowledgments
This work is financially supported by the National
Natural Science Foundation of China under Grant No. 12574298.
We acknowledge the support of the Supercomputing Center of Wuhan University.

\appendix
\section{Fourier transform form of $K$ on other lattices}
\label{append:defofmag}
As representative examples, we present the Fourier transform form of the kinetic matrix for the triangular and honeycomb lattices. These are among the most widely realized non-square optical lattice geometries and demonstrate that the present formalism can be readily extended beyond the kagome lattice.
\subsection{triangular lattice}\label{subsec:tri}
The spinless kinetic matrix $K$ on the triangular lattice can be expressed as
\begin{equation}\label{eq:tri_kron_plus}
  K=I\otimes K_s+K_s\otimes I+K_s\otimes K_s+ \text{h.c.}
\end{equation}
$K$ is Fourier transformed into the diagonal matrix $\Lambda^{\{1\cross 1\}}_{\text{tri}}$:
\begin{align}\label{eq:tri_k_kron_k}
  K
  =&\qty(F\otimes F)[I\otimes\Lambda_s+\Lambda_s\otimes I+\Lambda_s\otimes \Lambda_s+ \text{h.c.}]\qty(F^{\dagger}\otimes F^{\dagger})\notag\\
  =&\qty(F\otimes F)\Lambda^{\{1\cross 1\}}_{\text{tri}}\qty(F^{\dagger}\otimes F^{\dagger}).
\end{align}

\subsection{honeycomb lattice}\label{subsec:honey}
The spinless kinetic matrix $K$ on the honeycomb lattice can be expressed as
\begin{equation}\label{eq:hon_kron_plus}
  K=\qty(I\otimes I+I\otimes K_s+K_s\otimes I)\otimes\delta+ \text{h.c.},
\end{equation}
where the 2-dimensional matrices $\delta$ encodes the intra-cell hopping, containing a single nonzero entry at $(\delta)_{2,1}$.

$K$ is Fourier transformed into the block-diagonal matrix $\Lambda^{\{2\cross 2\}}_{\text{hon}}$:
\begin{align}
  K=&\qty(F\otimes F\otimes I_2)\qty[\qty(I\otimes I+I\otimes \Lambda_s+\Lambda_s\otimes I)\otimes \delta+ \text{h.c.}]\notag\\
  &\times \qty(F^{\dagger}\otimes F^{\dagger}\otimes I_2)\notag\\
  =&\qty(F\otimes F\otimes I_2)\Lambda^{\{2\cross 2\}}_{\text{hon}}\qty(F^{\dagger}\otimes F^{\dagger}\otimes I_2).
\end{align}
%

\end{document}